\documentclass[english]{article}

\usepackage{algorithmic}
\usepackage{algorithm}
\usepackage{amsfonts}
\usepackage{amsmath}
\usepackage{amssymb}
\usepackage{amsthm,wrapfig}
\usepackage{array}
\usepackage{arydshln}
\usepackage{bm}
\usepackage{cite}
\usepackage{color}
\usepackage{comment}
\usepackage{dsfont}
\usepackage{enumitem}
\usepackage{float}
\usepackage[T1]{fontenc}
\usepackage{geometry}
\usepackage{graphicx}
\usepackage{grffile}
\usepackage{hyperref}\hypersetup{colorlinks,linkcolor=blue,anchorcolor=blue,citecolor=blue}
\usepackage[latin9]{inputenc}
\usepackage{letltxmacro}
\usepackage{mathrsfs}
\usepackage{mathtools}
\usepackage{multirow}
\usepackage{nicefrac}
\usepackage{subcaption}
\usepackage{tablefootnote}
\usepackage{verbatim}

\usepackage{todonotes}

\usepackage{booktabs}

\allowdisplaybreaks

\newcommand{\bp}{\bm{p}}
\newcommand{\bq}{\bm{q}}

\newcommand{\bx}{\bm{x}}

\newcommand{\bB}{\bm{B}}

\newcommand{\bI}{\bm{I}}

\newcommand{\bS}{\bm{S}}

\newcommand{\bepsilon}{\bm{\epsilon}}

\newcommand{\cA}{\mathcal{A}}

\newcommand{\cL}{\mathcal{L}}
\newcommand{\cM}{\mathcal{M}}
\newcommand{\cN}{\mathcal{N}}

\newcommand{\cP}{\mathcal{P}}

\newcommand{\bcE}{\bm{\mathcal{E}}}

\newcommand{\bcX}{\bm{\mathcal{X}}}
\newcommand{\bcY}{\bm{\mathcal{Y}}}

\newcommand{\RR}{\mathbb{R}}

\newcommand{\zero}{\bm{0}}

\allowdisplaybreaks
\makeatletter

\theoremstyle{plain} 

\theoremstyle{definition}
 
\theoremstyle{remark}

\definecolor{tian}{RGB}{0,150,0}
\definecolor{cm}{RGB}{250,0,200}
\definecolor{yc}{RGB}{255,0,0}
\definecolor{hd}{RGB}{0,180,200}

\definecolor{edits}{RGB}{250,100,100}

\newcommand{\obs}{\bcY}

\begin{document}
\title{Multimodal Diffusion to Mutually Enhance\\Polarized Light and Low Resolution EBSD Data}

\author
 {
 	Harry Dong \thanks{Email: \texttt{harryd@andrew.cmu.edu}} \thanks{Department of Electrical and Computer Engineering, Carnegie Mellon University, USA} \\
	CMU 
	\and
    Timofey Efimov \footnotemark[2] \\
    CMU
    \and
    Megna Shah \thanks{Materials and Manufacturing Directorate, Air Force Research Laboratory, USA} \\
    AFRL
    \and
    Jeff Simmons \footnotemark[3] \\
    AFRL
    \and
    Sean Donegan \footnotemark[3] \\
    AFRL
    \and
    Marc De Graef \thanks{Department of Materials Science and Engineering, Carnegie Mellon University, USA}\\
    CMU
    \and
    Yuejie Chi \thanks{Department of Statistics and Data Science, Yale University, USA} \\
    Yale University
 }

\date{\today}

\setcounter{tocdepth}{2}
\maketitle

\begin{abstract}
In spite of the utility of 3-D electron back-scattered diffraction (EBSD) microscopy, the data collection process can be time-consuming with serial-sectioning. Hence, it is natural to look at other modalities, such as polarized light (PL) data, to accelerate EBSD data collection, supplemented with shared information. Complementarily, features in chaotic PL data could even be enriched with a handful of EBSD measurements. To inherently learn the complex dynamics between EBSD and PL to solve these inverse problems, we use an unconditional multimodal diffusion model, motivated by progress in diffusion models for inverse problems. Although trained solely on synthetic data once, our model has strong generalizable capabilities on real data which can be low-resolution, noisy, corrupted, and misregistered. With inference-time scaling, we show gains in performance on a variety of objectives including grain boundary prediction, super-resolution, and denoising. With our model, we demonstrate that there is little difference from full resolution performance with only 25\% (1/4 the resolution) of EBSD data and corrupted PL data.
\end{abstract}



\section{Introduction}
\label{sec:intro}

Electron BackScatter Diffraction (EBSD) is a workhorse modality for modern microstructure characterization because of the richness of the information it provides. Since each pixel in the output orientation map is a result of a machine indexed full diffraction pattern, this makes serial data acquisition with EBSD a potential bottleneck in 3-D serial sectioning experiments, and much attention has been dedicated to improving the throughput of the technique, both in the form of faster and more sensitive detectors \cite{degraef2026h} and better indexing algorithms \cite{degraef2026c}. By contrast, polarized light (PL) microscopy is very fast, since it acquires an entire image in parallel, but the technique suffers from the fact that it can only be applied to materials with a non-cubic crystal structure; in optically uniaxial materials like hexagonal close-packed Ti, only the orientation of the crystallographic $c$-axis can be determined, leaving the orientation in the plane normal to this axis as a degree of freedom.  For optically bi-axial materials this limitation does not apply. Nevertheless, it is tempting to consider a data fusion approach between minimal data acquisition with EBSD and the balance with PL microscopy in order to improve the overall throughput of microstructure characterization. This work presents a novel algorithm using generative artificial intelligence (AI), which applies inverse solvers with diffusion to recover features of an EBSD map (i.e., a corresponding field of orientation values at every pixel), from an input polarized light image. An EBSD map can be useful both for the explicit orientations of each grain in a polycrystalline material and for the simpler problem of just finding the boundaries of each grain. Both of these issues can be addressed from the results of this work.

\begin{figure}[t]
\begin{center}
\includegraphics[width=0.8\columnwidth]{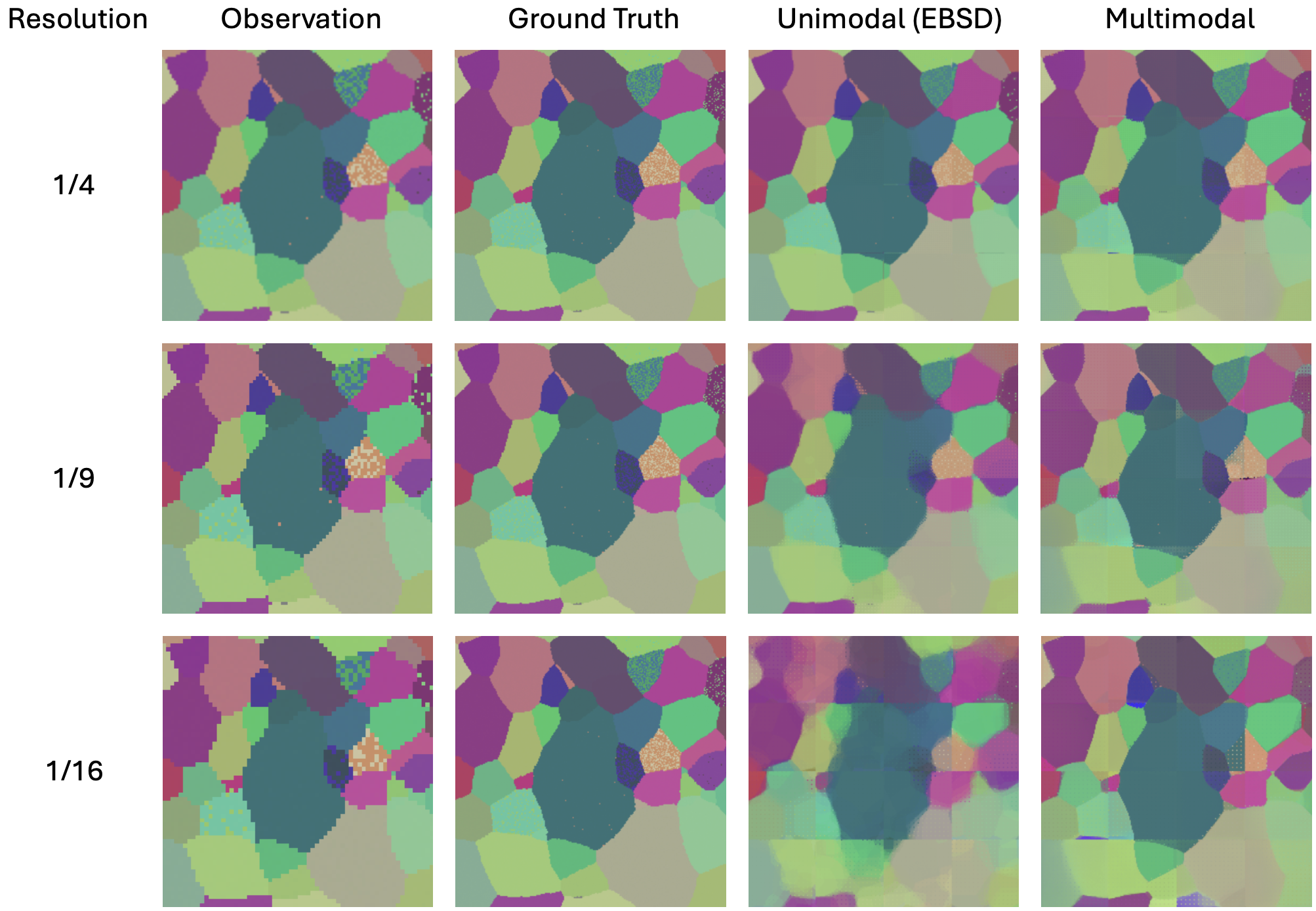}
\caption{Super-resolution using diffusion models. The multimodal (EBSD and PL) model is more robust as the observed resolution diminishes. Best viewed zoomed in.}  
\label{fig:sr_examples}
\end{center}
\end{figure}

Generally, \textit{we aim to use fewer EBSD measurements, which are relatively more expensive to collect, in conjunction with a greater quantity of cheaper PL measurements to accomplish downstream objectives that can benefit more from both modalities than either alone}. 
As we shall see, this is deeply rooted as an inverse problem. More concretely, the aim of solving inverse problems is to recover some underlying ground truth $\bcX^\star$ from observation $\bcY$ with the following relationship: 
\begin{align}\label{eq:forward}
    \obs = f(\bcX^\star) + \bcE, 
\end{align}
where $f$ is known as the forward model and $\bcE$ is measurement noise. Relating EBSD data as $\bcX^\star$ and PL data as $\bcY$ in this manner means $f = f_\text{E$\mapsto$P}$ is a complicated nonlinear and discontinuous function. To make matters worse, the inverse $f_\text{E$\mapsto$P}^{-1}$ is ill-posed since there is information in EBSD that cannot be captured by PL \cite{jin2018correlation,jin2020c}. However, \textit{we hypothesize at the slice level, spatial and macroscopic features in PL data can contain emergent information that is missing in pointwise low quality EBSD observations.} 

Diffusion models \cite{ddpm,song2020score} have been established as a highly expressive generative model to capture complex information in the data distribution in recent years.
Extensively tested on natural image generation \cite{peebles2023scalable, ramesh2022hierarchical, rombach2022high}, diffusion models have also shown efficacy as priors to solve inverse problems for natural \cite{chung2022diffusion,lugmayr2022repaint,xu2024provably}, medical \cite{lyu2209conversion, islam2023improving, webber2024diffusion}, and microscopy images \cite{efimov2025leveraging, saguy2025microtubule, efimov2026cross}. Since diffusion models are learned from the data of interest directly, they are capable of capturing characteristic that are difficult for pre-determined, hand-crafted priors such as sparsity \cite{baraniuk2007compressive, candes2006robust} or total variation \cite{rudin1992nonlinear}.
Diffusion models also generate images from randomly initialized noise, and repeated inference from different initializations with an inverse solver can lead to different output reconstructions of $\bcX^\star$ from the same input observation $\bcY$. 
Hence, \textit{inverse solvers with diffusion models have the advantage that instead of producing one reconstruction per observation, they can capture distributions of possible reconstructions via repeated inference}. Akin to Monte Carlo sampling from the posterior distribution, a set of output reconstructions provides measures of uncertainty, which are richer representations than a single predicted reconstruction. 


Though very successful in the domain of natural images, several technical challenges arise when we think about diffusion in our setting. One major challenge is that the forward model $f = f_\text{E$\mapsto$P}$ and its gradients are complex nonlinear functions, limiting the use of many existing diffusion-based inverse problem solvers. To correctly guide the reconstruction with multimodal information, we reformulate the original problem of reconstructing just one modality from another into joint multimodal reconstruction via a multimodal diffusion model, following Efimov et al. \cite {efimov2025leveraging}. By jointly learning the probability distribution of all modalities with a multimodal diffusion model, the new forward model becomes just a linear masking operator, $f_\text{EP}$. Consequently, the new reconstruction or inverse problem also becomes linear, bypassing the need to explicitly have a forward model between the two modalities. To illustrate the difference between unimodal and multimodal diffusion, let us adapt \eqref{eq:forward} for each case. Define perfect EBSD measurements, observed EBSD measurements, and EBSD measurement noise to be $\bcX^\star_\text{E}$, $\obs_\text{E}$, and $\bcE_\text{E}$ respectively, and similarly for PL ($\bcX^\star_\text{P}$, $\obs_\text{P}$, and $\bcE_\text{P}$). The forward model in the unimodal case is
\begin{align}\label{eq:uni_forward}
    \obs_\text{P} &= f_\text{E$\mapsto$P}(\bcX^\star_\text{E}) + \bcE_\text{P}.
\end{align}
The multimodal case concatenates the modalities into $\bcX^\star_\text{EP} = [\bcX^\star_\text{E} \quad \bcX^\star_\text{P}]$ for the target, $\obs_\text{EP} = [\obs_\text{E} \quad \obs_\text{P}]$ for the measurement, and $\bcE_\text{EP} = [\bcE_\text{E} \quad \bcE_\text{P}]$ for the noise. Hence, the new multimodal forward model becomes
\begin{align}\label{eq:multi_forward}
    \obs_\text{EP} &= f_\text{EP}(\bcX^\star_\text{EP}) + \bcE_\text{EP},
\end{align}
where $f_\text{EP}$ is a masking operation such that we only partially observe EBSD images but fully observe PL images. In practice, $f_\text{EP}$ can involve intentional choices like subsampling EBSD data to be at a lower resolution for faster data collection. By learning an unconditional multimodal diffusion model over both modalities, we implicitly capture the relationship between EBSD and PL data. 
\textit{Moreover, by learning an unconditional multimodal diffusion model as a joint prior on both modalities, the synthesis of information between EBSD and PL data can capture more holistic representations of the material than either alone.}

The second challenge is the lack of real EBSD data. This is a major issue as diffusion models are typically trained on a large corpus of data. To overcome this, we enlist the help of diverse synthetic data created by DREAM.3D \cite{groeber2014dream} to train our models. In fact, without touching any real data during training, we are able to see significant gains in performance on real data. In summary, we propose a general method for various reconstruction tasks involving EBSD and PL data which has the following benefits:

\begin{enumerate}
    \item \textbf{Transfer to Real Data:} Although we only use synthetic EBSD and PL data during training due to scarcity of real data, our method still has strong performance when deployed on real data without any additional training. 
    \item \textbf{Parallel Inference Scaling:} By sampling multiple reconstructions of $\bcX^\star$ from a single observation $\bcY$, we are able to obtain both a distribution of reconstructions and after aggregating, a higher quality objective-specific prediction than a single inference call.
    \item \textbf{Task Generalizability:} Although we focus primarily on \textit{boundary prediction} in this paper, we show that the same diffusion model and inference pipeline with minor post-processing adjustments can be used for other tasks such as \textit{EBSD super-resolution} and \textit{PL denoising}.
\end{enumerate}

Comprehensive experiments and evaluations show strong efficacy of parallel inference scaling for inverse solvers with unconditional multimodal diffusion for a variety of scenarios. For the rest of this section, we briefly cover relevant background and related works. Afterwards, we outline our training and inference pipelines in Section~\ref{sec:method}. In Section~\ref{sec:experiments}, we demonstrate our method's strong performance across a variety of settings where EBSD and PL data fusion is helpful such as boundary prediction, super-resolution, and denoising. Finally, we end with some concluding thoughts in Section~\ref{sec:conclusion}.

Throughout this paper, we denote scalars, vectors, matrices, and tensors with unstylized (e.g., $t$), bold lowercase (e.g., $\bx$), bold uppercase (e.g., $\bB$), and bold calligraphic uppercase letters (e.g., $\bcX$), respectively. When indexing into these arrays, we use brackets with subscripted indices (e.g., $[\bcX]_{i, j}$). Other less frequently used objects such as functions and sets will be defined ad hoc.

\subsection{Background}
\label{sec:background}

Our work builds upon ideas from machine learning which we briefly describe here. More specifically, we provide brief overviews on diffusion models, inverse problems, and inference scaling for generative models. 

\subsubsection{Diffusion Models}

At a high level, diffusion models \cite{ddpm,song2020score} map pure noise to some data distribution of interest with the probability density function $p(\bx)$. By sampling initializations, one can obtain generated data that ideally fall in the data distribution. The diffusion process is split into the forward process (not to be confused with the forward model $f$) and the backward process. Here, we let data points be vectors for simplicity, but the same formulations naturally carry over to matrices and tensors.

The forward process incrementally adds Gaussian noise to a sample $\bx_0 \in \RR^D$ from the data distribution $p_{\text{data}}$ such that
\begin{align}
    \bx_t = \sqrt{1 - \beta_t} \bx_{t-1} + \sqrt{\beta_t} \bepsilon_t,
\end{align}
for $1 \leq t \leq T$, noise vectors $\bepsilon_t$ drawn i.i.d. from $\cN(\zero, \bI_D)$, and learning rates $\beta_t \in (0, 1)$ that determine the amount of noise added at each step. In a single step, this reduces to 
\begin{align}
    \bx_t = \sqrt{\overline{\alpha}_t}\,\bx_0 + \sqrt{1-\overline{\alpha}_t}\,\bepsilon, \quad \bepsilon \sim \mathcal{N}(\mathbf{0}, \bI_D),
\end{align}
where 
\begin{align}
    \alpha_t := 1 - \beta_t, \quad \overline{\alpha}_t := \prod_{k=1}^{t} \alpha_k.
\end{align}
Furthermore, for some $\bx \in \RR^D$, the score function is defined as
\begin{align}
    s_t^\star(\bx) = \nabla_{\bx} \log p_t(\bx),
\end{align}
where $p_{t}$ is the probability density function of $\bx_t$. Intuitively, the score is the gradient that points in the direction of higher probability images. For the case of denoising from a Gaussian distribution, the score is the direction towards a less noisy image.

The backward process incrementally removes noise, starting with a Gaussian sample $\bx_T^{\text{rev}}$:
\begin{align}\label{eq:backward_chain}
    \bx_T^{\text{rev}} \rightarrow \bx_{T-1}^{\text{rev}}    \rightarrow \cdots \rightarrow \bx_0^{\text{rev}}.
\end{align}
In practice, we do not have access to $s_t^\star(\bx)$, and learn it by connecting score matching to denoising \cite{vincent2011connection}. In summary, a denoising network, denoted as $\cM_\theta(\bx, t)$, directly predicts the noise and thus implicitly approximates the score via
\begin{align}
    s_\theta(\bx_t, t) = -\frac{\cM_\theta(\bx_t,t)}{\sqrt{1-\overline{\alpha}_t}}.
\end{align}
Substituting into the score matching objective and simplifying yields the \emph{denoising loss}~\cite{ddpm}:
\begin{align}\label{eq:denoising_loss}
    \mathcal{L}(\theta) 
    = \mathbb{E}_{t,\, \bx_0,\, \bepsilon}\!\left[
        \left\|\bepsilon - \cM_\theta\!\left(\sqrt{\overline{\alpha}_t}\,\bx_0 + \sqrt{1-\overline{\alpha}_t}\,\bepsilon,\; t\right)\right\|^2
    \right],
\end{align}
where the expectation is over $t \sim \mathrm{Uniform}\{1,\dots,T\}$, $\bx_0 \sim p_{\mathrm{data}}$, and $\bepsilon \sim \mathcal{N}(\mathbf{0},\bI_D)$.

\subsubsection{Diffusion for Inverse Problems}

Unconditional diffusion models provide a powerful prior on the distribution of the data and can be effectively utilized to solve inverse problems without any task-specific training. The inverse solver in that case is then formulated as posterior sampling, where we try to enforce both the data consistency and prior distribution constraints. Recalling the forward model formulation from \eqref{eq:forward}, we want to sample $\hat{\bcX}$ from
\begin{align}
   p(\hat{\bcX} \mid \obs) \propto p^\star(\hat{\bcX}) p(\obs  \mid \bcX^\star = \hat{\bcX}),
\end{align}
where $p^\star$ is the prior distribution of $\bcX^\star$ (captured by the diffusion model), and $p(\obs  \mid \bcX^\star = \hat{\bcX})$ is the likelihood of observing $\bcY$ assuming the ground truth $\bcX^\star = \hat{\bcX}$ (assumed to be known). 
In turn, we sample from both prior and likelihood distributions separately to obtain posterior samples.

Several posterior sampling algorithms have been proposed with provable consistency guarantees \cite{xu2024provably, fpsscm, cardoso2023monte}. In this paper, we use one of them, FPS-SMC \cite{fpsscm}, a particle filtering algorithm for linear inverse problems. Recall that even though the forward model $f$ between modalities is not linear in general, the multimodal formulation recasts this problem as inpainting, which is linear \cite{efimov2025leveraging}.

\subsubsection{Parallel Inference Scaling: Multiple Generations to One Input}

For a single input, many classical and modern methods have explored producing and combining multiple intermediate outputs to obtain a refined output. First classically, ensemble methods like random forests and bagging train multiple sub-models such that during inference, each sub-model produces an output which is merged with others (e.g., averaging or voting) to produce the final output which is often more reliable \cite{breiman1996bagging, breiman2001random}. With training-intensive neural network architectures like large language models (LLMs) \cite{vaswani2017attention} and diffusion models, training multiple models is computationally and economically impractical. However, since some of these models are stochastic during inference, multiple outputs can be sampled from the same model and same input. This has led to significantly improved output quality for LLMs \cite{wang2022self, brown2024large, snell2024scaling, dong2025generalized} and diffusion models \cite{verdun2025soft, jain2025diffusion, zhang2025t}. In this paper, we take the same intuition to sample and combine multiple independent reconstructions from the same inverse solver, diffusion model, and observation to improve the final reconstruction quality and stability.

``Parallel inference scaling'' is modern jargon common in LLM and generative AI literature, but this idea is rooted in concepts that have existed before, namely posterior sampling. LLMs and diffusion models are stochastic generators and thus can be abstracted away as models of distributions, so vanilla parallel inference scaling is direct Monte Carlo sampling from these distributions. Concretely in our setting, parallel inference scaling entails repeated independent sampling from the posterior distribution, $p(\hat{\bcX} \mid \bcY)$ (i.e., sampling multiple reconstructions with the same observation $\bcY$ and same inverse solver but with different diffusion noise initializations described in \eqref{eq:backward_chain}). Analogously, refinements such as importance sampling \cite{faria2025sample, yao2025your} and rejection sampling \cite{leviathan2023fast, verine2024optimal, chen2026jackpot} also exist for these generative models. Inference scaling is a broader class of methods that allocates more computational resources during inference which contrasts with other aspects of scaling in deep learning like training data and model size. For this paper, we opt to use the term ``parallel inference scaling'' to stay consistent with the generative AI literature.

\section{Materials and Methods}
\label{sec:method}

\begin{figure}[t]
\begin{center}
\includegraphics[width=0.65\columnwidth]{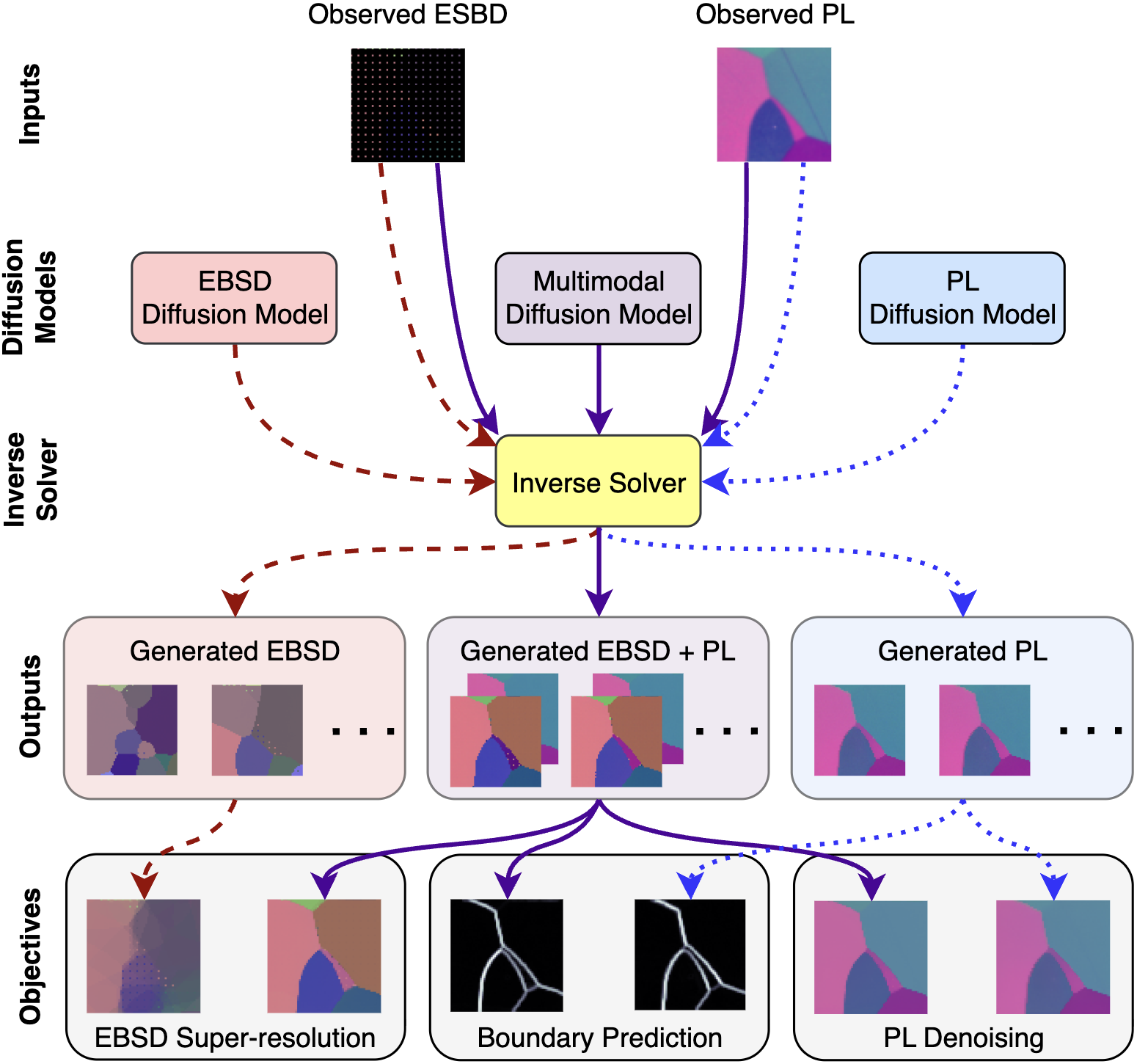}
\caption{Inference algorithm. The dashed red arrows, dotted blue arrows, and solid purple arrows indicate the inference process when the prior distribution is modeled by an EBSD-only, PL-only, or multimodal diffusion model, respectively. All diffusion models are unconditional generative models. Regardless of the diffusion model, given observations of EBSD and/or PL data, we sample multiple outputs (reconstructions $\{\widehat{\bcX}_i\}^N_{i=1}$) per input (observation $\bcY$) using the same inverse solver. These outputs are then processed in downstream objectives like super-resolution, boundary prediction, and denoising. Unimodal diffusion models are restrictive in that they only be applied to a narrow set of tasks that rely on the modality they were trained on. The multimodal diffusion model mixes information from and jointly generates EBSD and PL data to be applicable to multimodal objectives and the union of tasks that these unimodal models are used for, often even surpassing unimodal performance. The resolution of the observed EBSD in this example is 1/16.}
\label{fig:algorithm}
\end{center}
\end{figure}


We train three models: \textbf{(1)} a multimodal diffusion model, $\cM_\text{EP}$, that generates EBSD and PL images; \textbf{(2)} a unimodal diffusion model, $\cM_\text{E}$, that generates EBSD images; \textbf{(3)} a unimodal diffusion model, $\cM_\text{P}$, that generates PL images. We omit $\theta$ as used in \eqref{eq:denoising_loss} for brevity.

\subsection{Data Formulation and Preprocessing} 
The input data takes multiple forms depending on the model. Define $H$, $W$, $D_\text{E}$, and $D_\text{P}$ to be the image height, image width, number of EBSD features, and number of PL features, respectively. Letting $\bcX^\star_\text{E} \in \RR^{H \times W \times D_\text{E}}$ and $\bcX^\star_\text{P} \in \RR^{H \times W \times D_\text{P}}$ be the true EBSD and PL data, respectively, we construct $\bcX^\star_\text{EP} = [\bcX^\star_\text{E} \quad \bcX^\star_\text{P}] \in \RR^{H \times W \times (D_\text{E} + D_\text{P})}$ to be the point-wise concatenation of these measurements. By \eqref{eq:multi_forward}, we observe $\obs_\text{E} \in \RR^{H \times W \times D_\text{E}}$, $\obs_\text{P} \in \RR^{H \times W \times D_\text{P}}$, and consequently, $\obs_\text{EP} = [\bcY_\text{E} \quad \bcY_\text{P}] \in \RR^{H \times W \times (D_\text{E} + D_\text{P})}$.
Since the models are trying to recover the true values from altered and incomplete ESBD data, we do not perform masking on PL data. Beyond white noise, other (perhaps systematic) real perturbations present in $\bcE_\text{E}$, $\bcE_\text{P}$, and $\bcE_\text{EP}$ can include corruptions, registration error, and subsampling (e.g., low-resolution). At inference, the inverse solver with diffusion model $\cM_\text{EP}$, $\cM_\text{E}$, or $\cM_\text{P}$ observes $\obs_\text{EP}$, $\obs_\text{E}$, or $\obs_\text{P}$, respectively.

Training data consist of synthetic EBSD and PL images while test data consist of real images. Synthetic EBSD volumes are generated from DREAM.3D \cite{groeber2014dream}, and corresponding synthetic PL data are computed with EMsoftOO \cite{emsoft} across $9$ equal $40^\circ$ rotations, standing in as the inter-modality forward model $f_{\text{E$\mapsto$P}}$. All PL data are compressed via principal component analysis, keeping the top $3$ principal components. All EBSD data are represented as cubochoric coordinates \cite{rocsca2014new, dong2023lightweight}. All values are then normalized to fall between $-1$ and $1$. The result is $3$ channels for EBSD and $3$ channels for PL for each voxel (i.e., $D_\text{E} = D_\text{P} = 3$). At each training batch, we sample slices of shape $H \times W \times D$, where $D$ depends on the modality of the model, from the volumes applied with random geometric data augmentation techniques like rotations and flips. The models are trained to unconditionally recover these samples.

\begin{figure}[t]
\begin{center}
\includegraphics[width=\columnwidth]{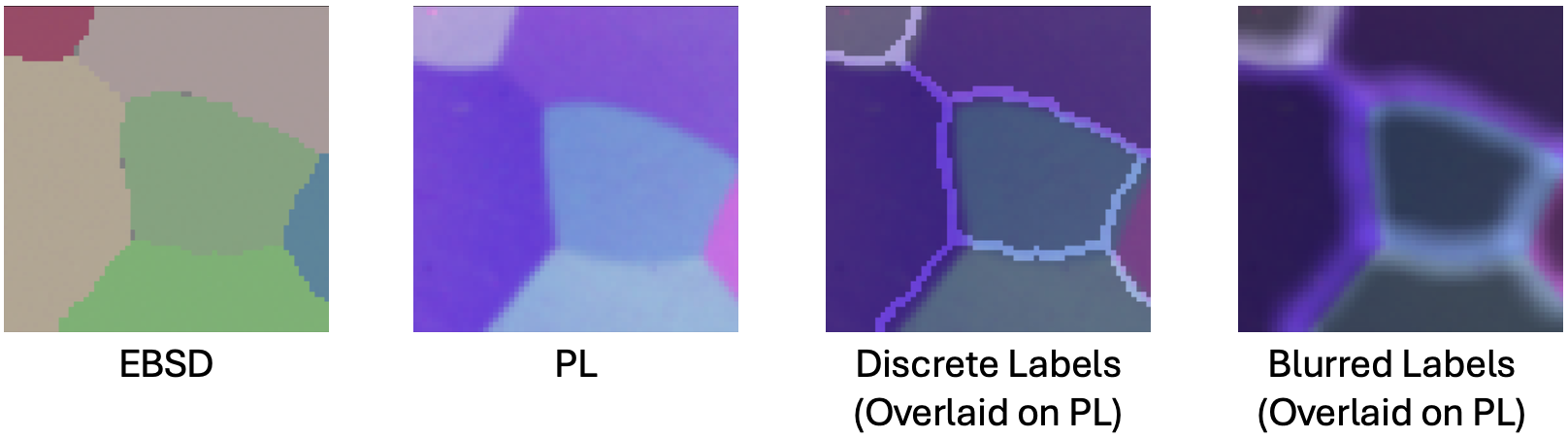}
\caption{Due to image registration errors, grains in EBSD data and PL data may have slight positional mismatches. We highlight this difference in the third image from the left where ground truth boundary labels $\bB^\star$ calculated from EBSD do line up with boundary pixels in PL data. Gaussian blurred ground truth boundaries $g(\bB^\star)$ overlaps with boundaries present in PL data, helping BCE loss be robust to slight spatial shifts in predictions. While this directly impacts boundary prediction, this can impact the generation process and evaluation of other objectives as well. }
\label{fig:labels}
\end{center}
\end{figure}

To evaluate, we test on real data from a Ti7 specimen. The sample was part of a serial sectioning experiment, but only a one section was used here. The sample was polished using the second-generation RoboMet.3D\textsuperscript{TM} starting with 3$\mu$ diamond slurry and Struers Largo pad, followed by a 3$\mu$ diamond slurry on an Allied Gold Label pad, followed by water on a cleaning pad, followed by 40nm colloidal silica on an Allied Final A pad and ending with water on a cleaning pad. EBSD on the sample was collected with a Tescan Vega SEM at 20 keV and a Bruker Quantax e-Flash 1000 EBSD detector at 3.03 micron pixel size. PL imaging was done using a Zeis Axiovert 200M Optical Microscope and the pixel resolution was 1.04 microns, with the stage rotated every $40^\circ$. Since EBSD maps are being used as a ground truth, they need to be corrected for distortions and registered to the PL image. The images were registered using the Simple ITK framework \cite{simpleITK} and a methodology previously described \cite{afrl_midas_2021}. The fixed image was one PL image, downsampled to the same resolution as the EBSD image. The moving image was a grayscale of the inverse pole figure-X map.  A bspline transform warped the collected EBSD data to match the PL data, using mutual information as a metric. While the registration significantly improves the pixel-to-pixel matchups between modalities, some registration error may still exist (Figure~
\ref{fig:labels}).

Note that EBSD and PL images in this paper have been shifted to aid visual intuition and should not be taken as actual values. Our quantitative evaluation metrics are based on the actual values.

\subsection{Architecture and Training} 
All diffusion models are a $6$-layer $143$M parameter transformer-based U-Nets \cite{efimov2025leveraging, ronneberger2015u}. We use a hidden dimension of $92$, which doubles with every downsampling operation and halves with every upsampling operation. The only architectural difference between multimodal and unimodal diffusion models is the channel dimension of the input and output layers. The height ($H$) and width ($W$) are set to $64$. We train the models on two textured $200 \times 200 \times 320$ volumes, taking random $64 \times 64$ slices for each sample. We use random slices from two $200 \times 200 \times 70$ volumes as validation. In total, we train for $30$K gradient steps with a learning rate of $5\mathrm{e}{-4}$ and batches of $128$ samples, evaluating at every $100$ steps to select the best checkpoint. The training loss is the empirical denoising loss described by \eqref{eq:denoising_loss}.

\subsection{Inference Pipeline} 
\label{sec:inference_pipeline}
We depict the general inference pipeline in Figure~\ref{fig:algorithm}. We use FPS-SMC \cite{dou2024diffusion}, a particle filtering generation algorithm for linear inverse problems with provable consistency guarantees. \textit{In all experiments, we observe the full resolution but noisy PL data and sparsely subsampled EBSD data.} To scale inference, we generate a set of $N$ reconstructions $\{\widehat{\bcX}_i\}^N_{i=1}$ from a single observation $\bcY$ where their exact shapes depend on the number of modalities. Note that each $\widehat{\bcX}_i$ can be generated in parallel. Although the $\widehat{\bcX}_i$'s are Monte Carlo samples that can provide information of the reconstruction distribution, it may also be necessary to aggregate them into one final output. This procedure varies depending on the objective. For the rest of this section, we provide three example settings that we explore in this paper and their aggregation methods, all of which perform post-processing on the same reconstruction set $\{\widehat{\bcX}_i\}^N_{i=1}$.

\subsubsection{Grain Boundary Prediction}

In this task, we want to make binary pixel-wise predictions on whether a pixel is on the boundary of a grain. Due to the sparse structure of EBSD and PL data \cite{dong2023deep}, grain boundaries capture uniquely critical information. A boundary pixel is defined to be any pixel that has a face-sharing neighbor from another grain. This information is present in real and synthetic EBSD data where each pixel contains an ID that indicates the grain it belongs in. Pixels with null IDs are ignored.

We use Sobel filtering \cite{sobel19683x3}, a common edge detection method, on each reconstructed PL image followed by 0-1 normalization to produce gradient magnitude maps $\{\bS_i \in [0,1]^{H \times W} \}^N_{i = 1}$. Taking the average gradient map, we get $\Bar{\bS} = \frac{1}{N} \sum_i \bS_i$, which roughly measures the confidence that a pixel is a boundary. From here, we can obtain discrete predictions of boundary locations $\bB \in \{0, 1\}^{H \times W}$ where boundary pixels are set to 1. While setting a fixed cutoff for all images is possible, we choose a more adaptive method to automatically identify a custom cutoff per observation. First, we sort the pixel values of $\Bar{\bS}$ in descending order and smooth the resulting curve with a Gaussian kernel. Then, we use the kneed algorithm \cite{satopaa2011finding} which uses estimated second-order information to identify the elbow of the curve. All pixels at or above this cutoff are predicted as boundaries. Since we use PL channels for Sobel filtering (generated boundaries derived from EBSD have greater variance), we compare the performance among the multimodal diffusion model $\cM_\text{EP}$, PL diffusion model $\cM_\text{P}$, and a Sobel filter directly applied to observed PL data.


\subsubsection{EBSD Super-resolution}

The goal of super-resolution is to enhance an observed image into a higher resolution image. This is well suited for EBSD where a lower resolution dramatically speeds up the time-intensive process of data collection. We formulate super-resolution as an inpainting problem. Observed pixels are scattered to the full image size in a uniform grid which is used as $\bcY$. The model recovers the remaining unobserved EBSD pixels. Then, a neural network learns to align the reconstructed EBSD values at the observed positions with observed EBSD values and applies the same transformation to all pixels, a simplified post-processing method of Arefeen et all. \cite {arefeen2024infusion}.

\begin{figure}[t]
\begin{center}
\includegraphics[width=\columnwidth]{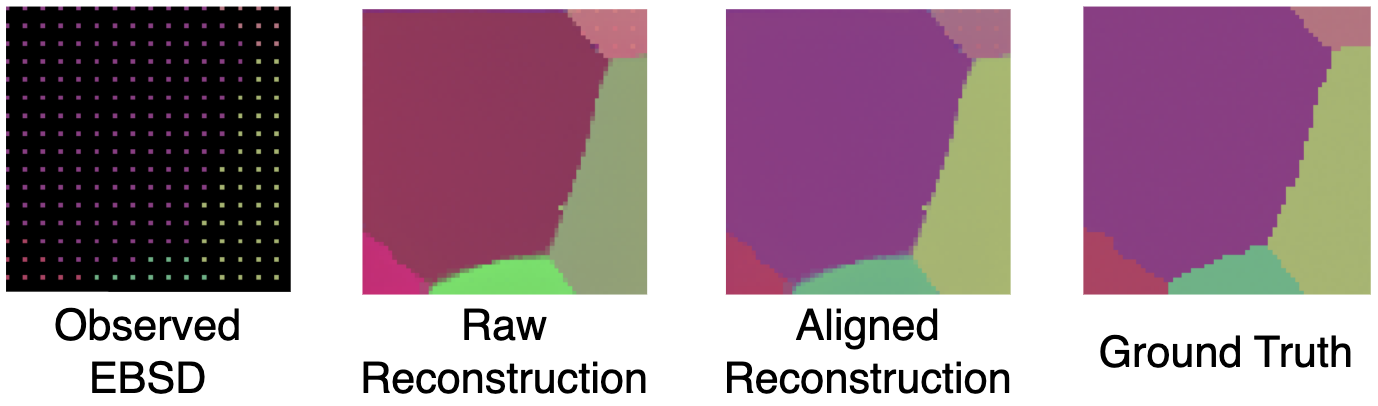}
\caption{Effect of post-reconstruction alignment for super-resolution. By training an alignment network, $\cA_\theta$, to learn a mapping to observed EBSD values from corresponding pixels of reconstructed EBSD values, we can reduce systematic error which better recovers the ground truth. This super-resolution example shows an observed EBSD image with a resolution of 1/16.}
\label{fig:sr_align}
\end{center}
\end{figure}

In more detail, we again start with the reconstruction set $\{\widehat{\bcX}_i\}^N_{i=1}$. First, we take the average to obtain $\Bar{\bcX} = \frac{1}{N} \sum_i \widehat{\bcX}_i$. Since EBSD data are comparatively more uniform within a grain than PL data (e.g., see Figure~\ref{fig:labels}), we use observed EBSD values to align reconstructed EBSD values at the same positions to reduce systematic error like minor shifts in values, as generated EBSD images are also fairly uniform within a grain. More concretely, let $\Omega$ be the set of observed coordinates in EBSD (i.e., nonempty coordinates in $\bcY_\text{E}$). For example, if we observe a resolution of 1/16, $\Omega$ would contain the grid coordinates of every 4th row and column. For each observation, we train small neural network, $\cA_\theta$, that minimizes the following pixel-wise mean squared error loss:
\begin{align}
    \cL_{\text{MSE}}(\theta) = \frac{1}{|\Omega|} \sum_{(i, j) \in \Omega} \| \cA_\theta([\Bar{\bcX}]_{i,j}) - [\bcY_\text{E}]_{i,j} \|^2_2.
\end{align}
Finally, we apply this same neural network to the all pixels in $\Bar{\bcX}$ to align every EBSD value for the final output $\cA_\theta(\Bar{\bcX}) \in \RR^{H \times W \times D_{\text{E}}}$. The effect of alignment is visualized in Figure~\ref{fig:sr_align}. We compare performance between the multimodal diffusion model $\cM_\text{EP}$ and EBSD diffusion model $\cM_\text{E}$. The PL model $\cM_\text{P}$ is not applicable here.

The alignment network $\cA_\theta$ is a $2$-layer neural network with a hidden dimension of $32$ separated with a GELU activation \cite{hendrycks2016gaussian}. Setting aside $20\%$ of $\Omega$ for validation, $\cA_\theta$ is trained with a learning rate of $0.02$ with Adam \cite{kingma2014adam} in batches of $16$ for $50$ epochs. Due to the minuscule network and data size, this takes less than $5$ seconds to train, a relatively insignificant cost.

\subsubsection{PL Denoising}

Denoising cleans an image of unwanted artifacts like noise and corruptions. Here, we adopt the straightforward process of averaging the PL values in the reconstruction set: $\Bar{\bcX}_{\text{P}} = \frac{1}{N} \sum_i \widehat{\bcX}_{\text{P},i}$. This is because the generation process will naturally produce a denoised image.

\begin{figure*}[!t]
\begin{center}
\includegraphics[width=\textwidth]{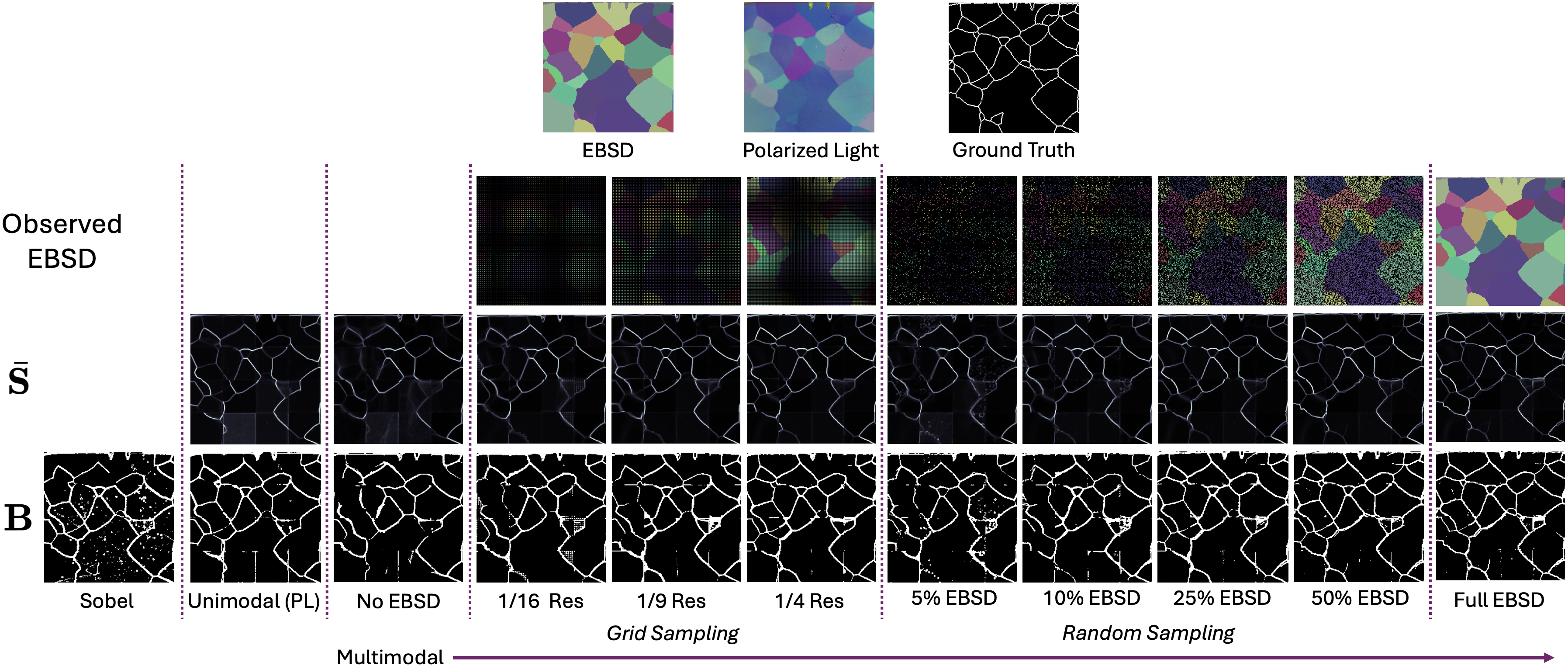}
\caption{Results of 16 $64 \times 64$ images stitched together for a continuous $256 \times 256$ image. From left to right, the top row shows the EBSD measurements, PL measurements, and ground truth boundaries $\bB^\star$ derived from EBSD. The second row from the top shows the observed EBSD, if relevant, in each setting. The third row shows the aggregated Sobel maps $\Bar{\bS}$ of diffusion-based methods with brighter colors indicating higher values. The bottom row shows the predicted boundaries $\bB$ of all methods. Best viewed zoomed in.}
\label{fig:stitched_edge_pred}
\end{center}
\end{figure*}

\begin{figure*}[!tb]
\begin{center}
\includegraphics[width=0.95\textwidth]{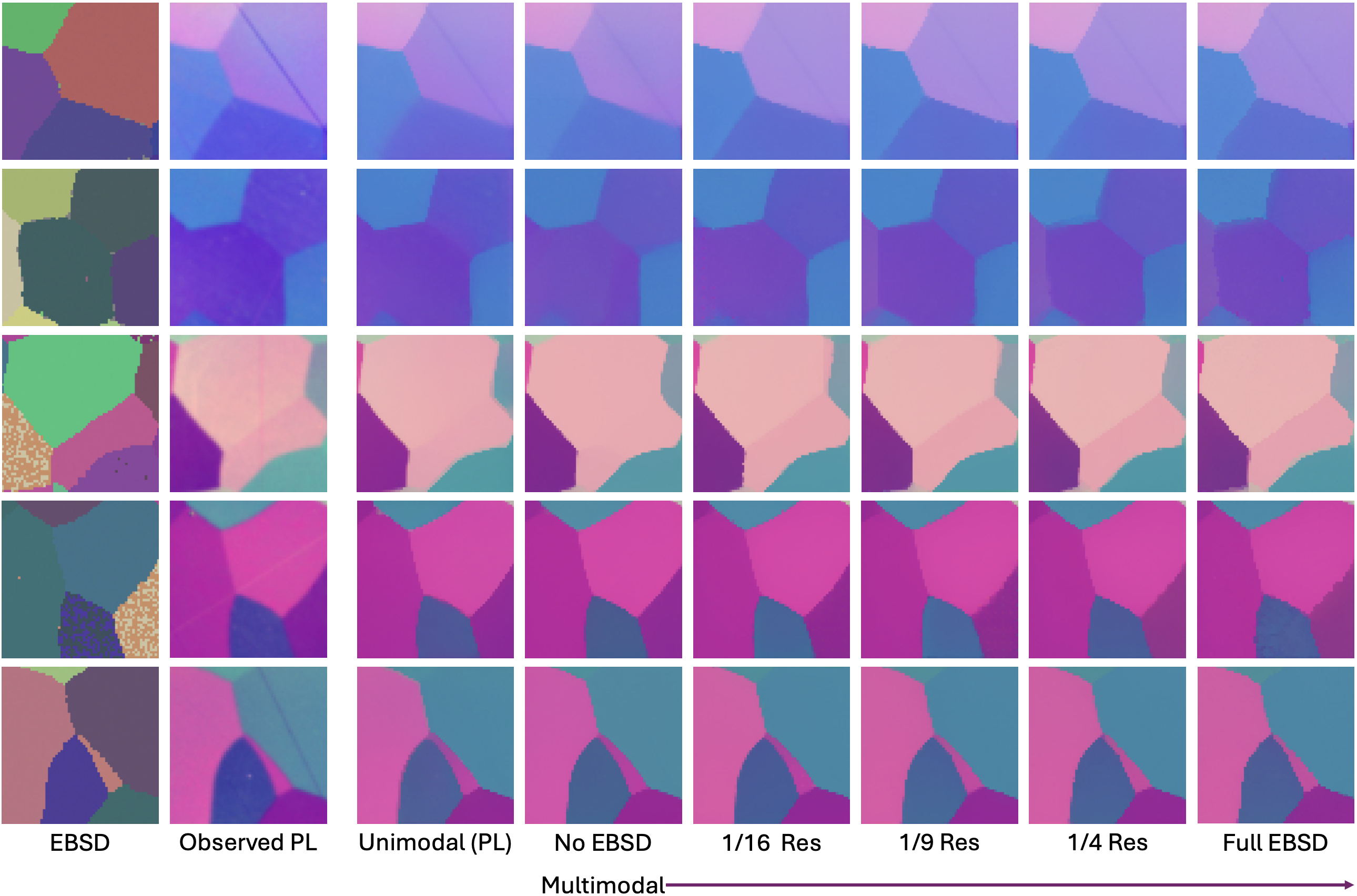}
\caption{Examples of PL denoising. The left two columns are the EBSD and PL data. Continuing to the right, the columns show the outputs from the unimodal model $\cM_{\text{P}}$ and multimodal model $\cM_{\text{EP}}$ (ranging from no EBSD to higher resolutions of observed EBSD).}
\label{fig:denoise}
\end{center}
\end{figure*}

\section{Results \& Discussion}
\label{sec:experiments}

Now, we showcase the strong performance our multimodal diffusion model across a variety of real data settings in Section~\ref{sec:performance}. Then in Section~\ref{sec:error}, we dive into sources of error and potential areas of improvement. 

\subsection{Performance}
\label{sec:performance}

Multimodal diffusion as a joint prior between EBSD and PL data demonstrates strong performance on multiple objectives including grain boundary prediction, super-resolution, and denoising. As described in Section~\ref{sec:inference_pipeline}, all objectives use the same reconstruction sets from all diffusion models and differ primarily in their post-processing procedures. Unless stated otherwise, we set the reconstruction set size $N = 10$ and repeat $10$ times to obtain error bars. 

\subsubsection{Grain Boundary Prediction}
\label{sec:exp_boundary}

Derived from EBSD grain IDs, let $\bB^\star \in \{0, 1\}^{H \times W}$ be the ground truth binary images with boundary pixels are set to 1. Since image registration is imperfect (see Figure~\ref{fig:labels}), point-wise metrics like accuracy and binary cross entropy (BCE) can accentuate errors where predictions and labels are slightly shifted from each other. Thus, we evaluate boundary prediction with two more positionally robust metrics: Chamfer distance and Gaussian blurred BCE. 

The Chamfer loss \cite{ravi2020pytorch3d} measures distance between point clouds $\cP = \{(\frac{i}{H}, \frac{j}{W}) | [\bB]_{i, j} = 1 \}$ and $\cP^\star = \{(\frac{i}{H}, \frac{j}{W}) | [\bB^\star]_{i, j} = 1 \}$, where $\bB$ is the predicted binary boundary image, defined as
\begin{align}
    C(\cP, \cP^\star) \coloneqq \underbrace{\frac{1}{|\cP|} \sum_{\bp \in \cP} \min_{\bq \in \cP^\star} \| \bp -  \bq \|^2_2}_{\overrightarrow{C}(\cP, \cP^\star)} + \underbrace{\frac{1}{|\cP^\star|} \sum_{\bq \in \cP^\star} \min_{\bp \in \cP} \| \bq -  \bp \|^2_2}_{\overrightarrow{C}(\cP^\star, \cP)}.
\end{align}
Here, $\overrightarrow{C}(\cP, \cP^\star)$ and $\overrightarrow{C}(\cP^\star, \cP)$ are asymmetric forward and backward Chamfer losses, respectively. The forward Chamfer loss intuitively penalizes overpredicting the number of boundary points (since $|\cP|$ depends on the predictions) and misclassifying distant points as a boundary, analogous to Type I error. Conversely, the backward Chamfer loss penalizes missing a boundaries, analogous to Type II error. 

As a metric to measure continuous predictions, we use $\cL_{\text{G-BCE}}$, BCE between aggregated normalized Sobel outputs, $\Bar{\bS}$, and Gaussian blurred ground truth boundaries, $g(\bB^\star)$. We opt to blur ground truth labels because of slight label mismatch between EBSD and PL images as depicted in Figure~\ref{fig:labels}. Blurring allows greater robustness to tolerate minor spatial discrepancies between predicted and target boundary pixels. More formally,
\begin{align*}
    \cL_{\text{G-BCE}}(\Bar{\bS}, \bB^\star)
    &= \frac{1}{HW}\sum_{i,j} [g(\bB^\star)]_{i, j} \log [\Bar{\bS}]_{i, j} + (1 - [g(\bB^\star)]_{i, j}) \log (1 - [\Bar{\bS}]_{i, j}).
\end{align*}
We set the standard deviation and radius of the Gaussian filter to be 3 and 5, respectively.

\begin{figure}[!t]
\begin{center}
\includegraphics[width=0.4\columnwidth]{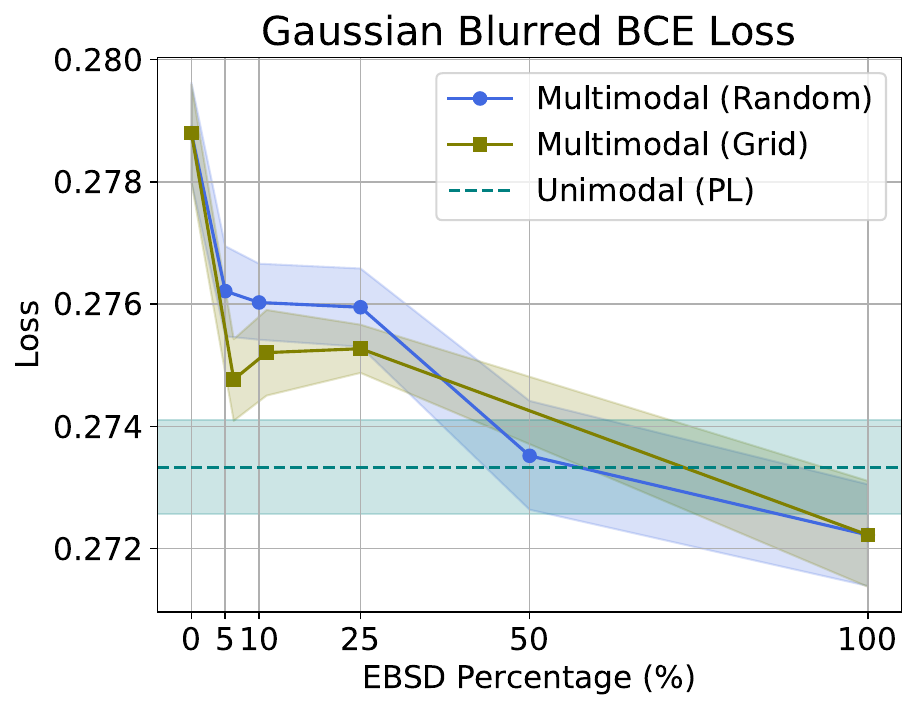}
\includegraphics[width=0.4\columnwidth]{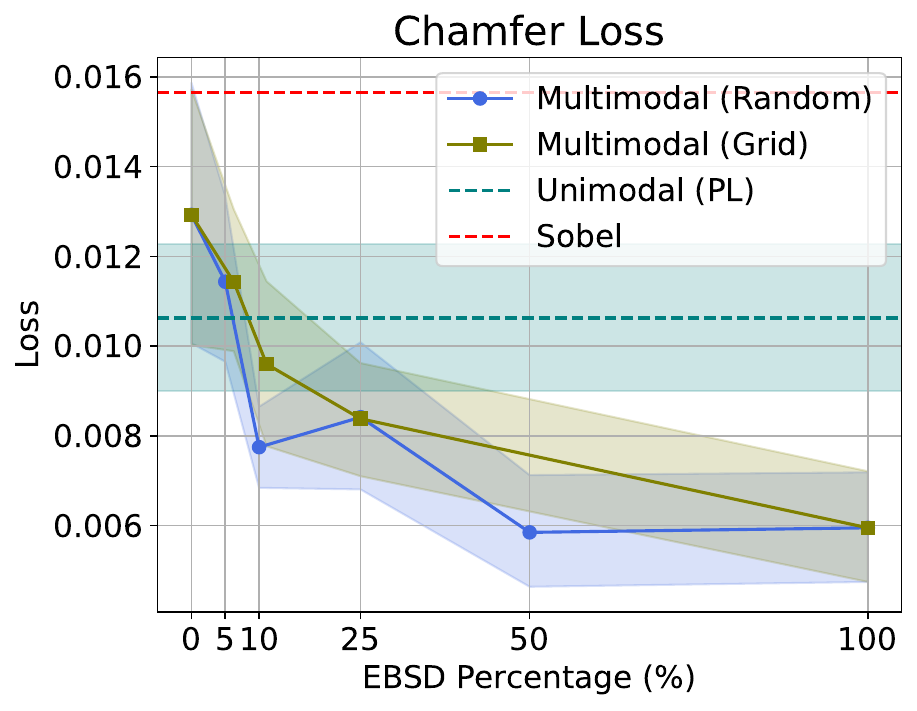}
\includegraphics[width=0.4\columnwidth]{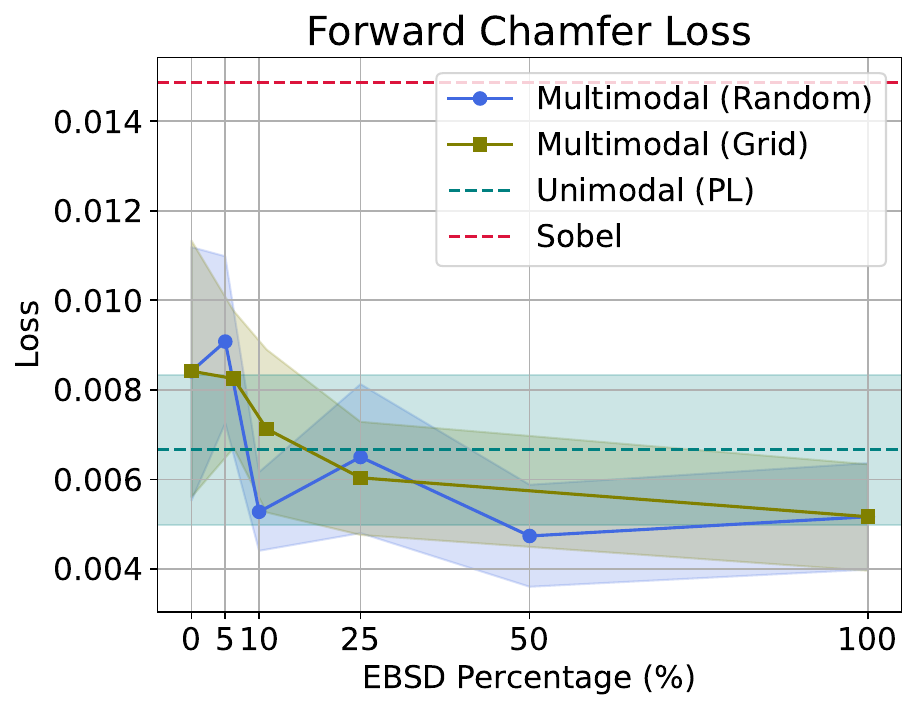}
\includegraphics[width=0.4\columnwidth]{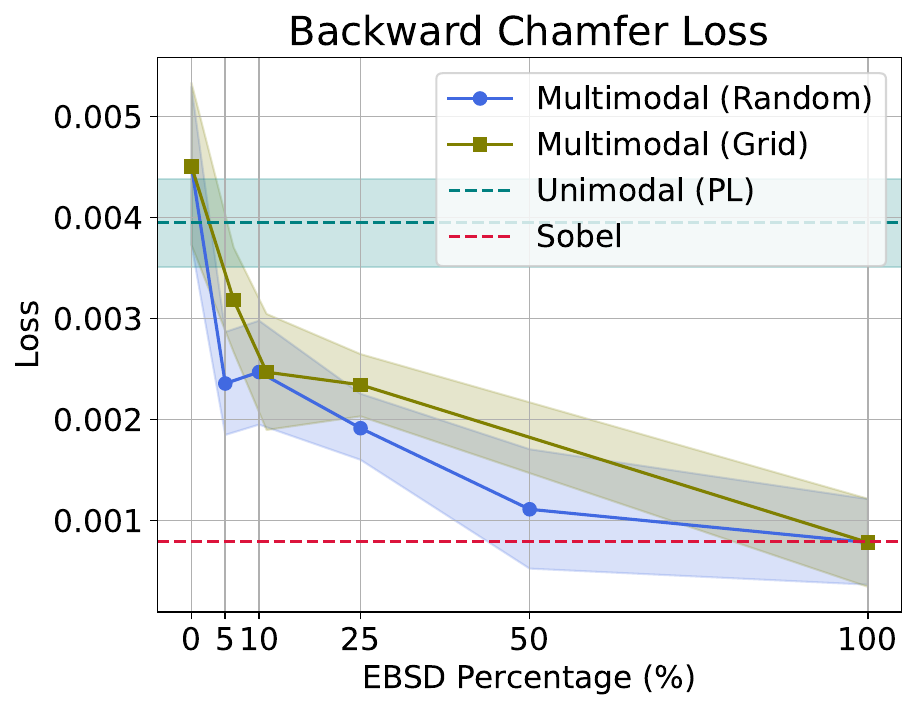}
\caption{Going clockwise starting from the top left, the Gaussian blurred BCE loss, Chamfer loss, backward Chamfer loss, and forward Chamfer loss of each model as more EBSD samples are observed. Multimodal results of randomly and uniformly subsampled (low-resolution) EBSD observations are shown in blue circles and dark green squares, respectively. The red dashed lines show the Sobel baseline performance. The Sobel baseline cannot be evaluated with Gaussian blurred BCE since it does not produce a probability distribution. Shaded regions illustrate two standard errors. Lower is better.}
\label{fig:boundary_metrics}
\end{center}
\end{figure}

\begin{figure}[!t]
\begin{center}
\includegraphics[width=0.4\columnwidth]{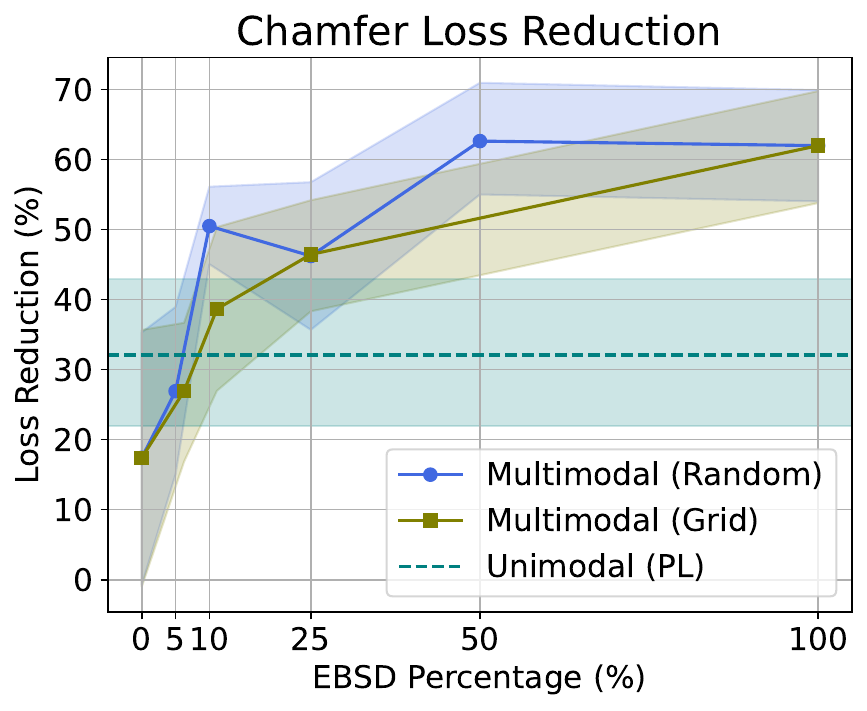}
\caption{Average percent reduction in Chamfer loss per sample compared to pure Sobel filtering. Multimodal results of randomly and uniformly subsampled (low-resolution) EBSD observations are shown in blue circles and dark green squares, respectively. The diffusion models all show greatly improved predictions. Shaded regions illustrate two standard errors. Higher is better.}
\label{fig:boundary_improvement}
\end{center}
\end{figure}

In Figure~\ref{fig:boundary_metrics}, we show comparisons between applying Sobel filtering on the PL data, the PL unimodal model $\cM_\text{P}$, and the multimodal model $\cM_\text{EP}$. We observe that multimodal diffusion naturally reduces all loss metrics as we observe more EBSD. Both diffusion models are overall much more accurate than simply applying a Sobel filter on PL observations, but the lone exception is the backward Chamfer loss. Even so, the forward Chamfer loss is significantly larger in magnitude and thus makes up most of the bidirectional Chamfer loss. This implies that the Sobel filtering baseline tends to make more Type I errors, which can be visually validated in Figure~\ref{fig:stitched_edge_pred} where many intra-grain pixels are predicted as boundaries. When we look at the percentage reduction in Chamfer loss in Figure~\ref{fig:boundary_improvement}, we see that the differences between the diffusion models and the Sobel filtering baseline are more pronounced. In general, the performance difference between random and grid (lower resolution) sampling observed EBSD pixels is not statistically significant. The slightly lower performance of multimodal diffusion in the complete absence of EBSD data is discussed in Section~\ref{sec:no_multi}.

\begin{figure}[t]
\begin{center}
\includegraphics[width=0.47\columnwidth]{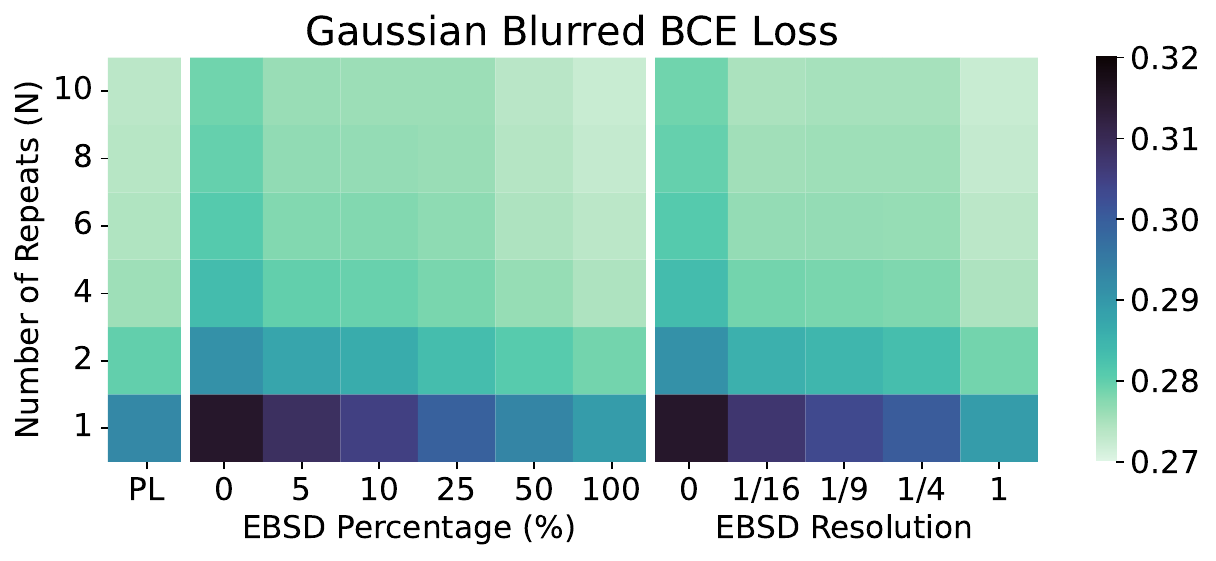}
\includegraphics[width=0.47\columnwidth]{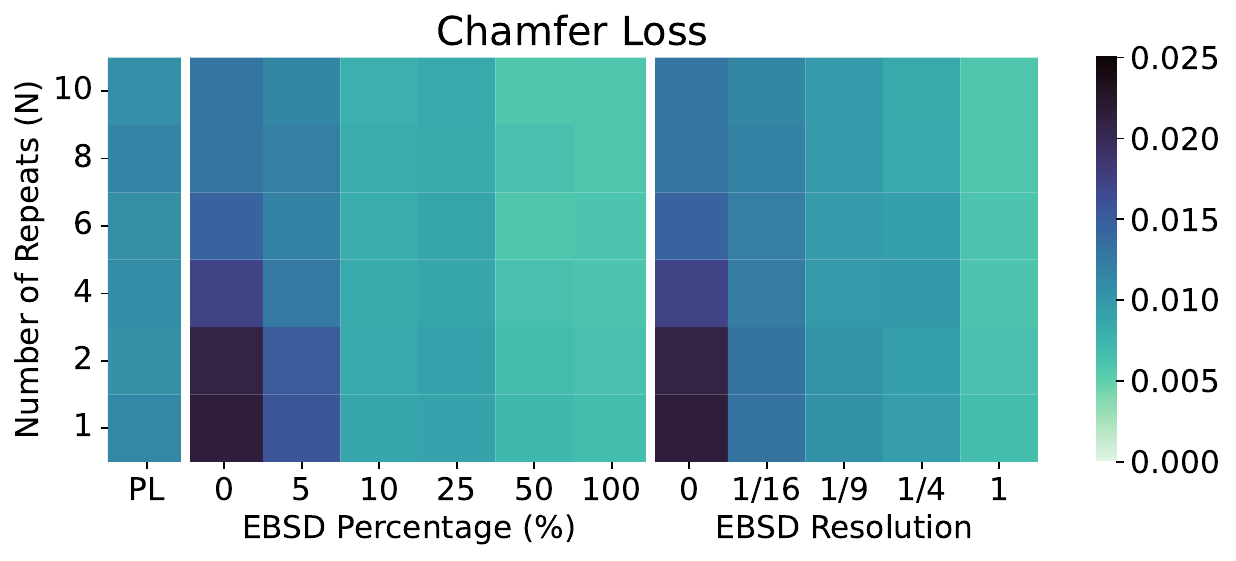}
\includegraphics[width=0.47\columnwidth]{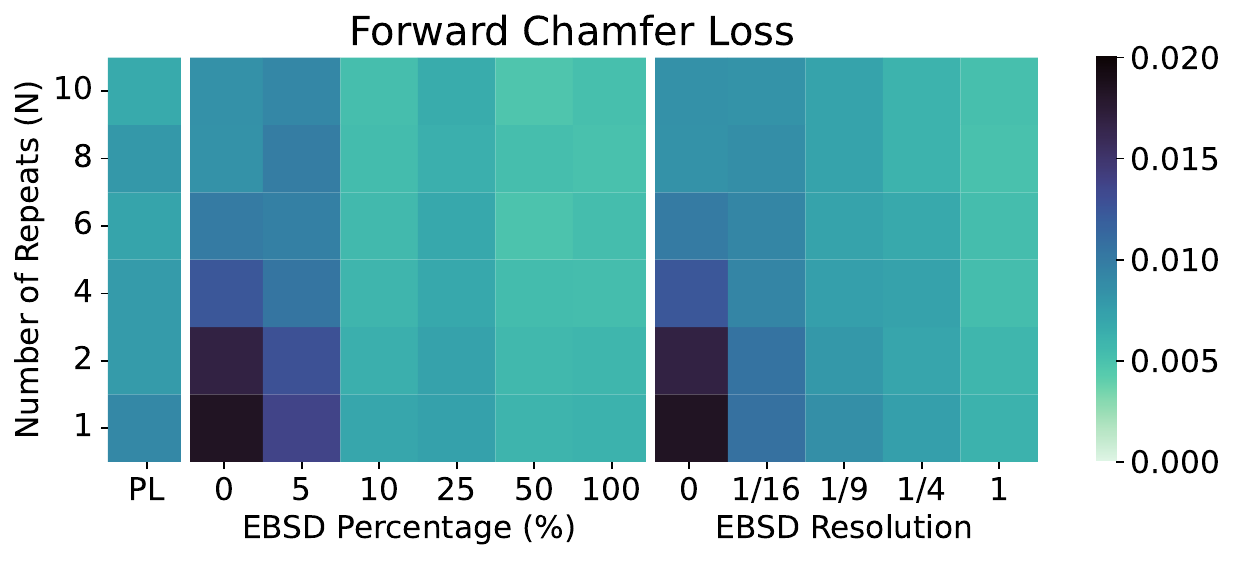}
\includegraphics[width=0.47\columnwidth]{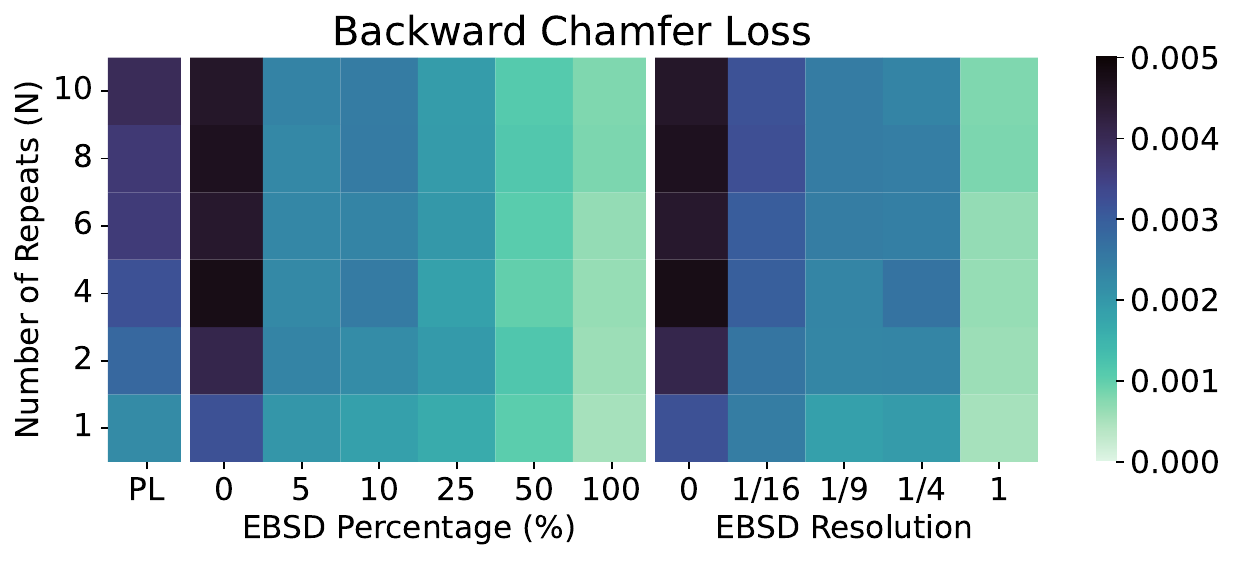}
\caption{Clockwise from the top left, the Gaussian blurred BCE loss, Chamfer loss, backward Chamfer loss, and forward Chamfer loss as we vary the number of repeats $N$ and observed EBSD percentage. From left to right, the blocks in each row show metrics from the unimodal model, multimodal model with randomly sampled EBSD observations, and multimodal model with low-resolution EBSD. We generally see improved performance as $N$ and the amount of observed EBSD increases. Note the difference in numerical ranges for each row. Lighter colors are better.}
\label{fig:loss_heatmaps}
\end{center}
\end{figure}

We also investigate the relationship between the amount of EBSD observed and $N$, the number of repeated generations per input. From Figure~\ref{fig:loss_heatmaps}, scaling $N$ generally helps both the unimodal and multimodal diffusion models. Hence, there is a tangible benefit to parallel inference scaling.

\subsubsection{Super-resolution Enhancement}

\begin{figure}[t]
\begin{center}
\includegraphics[height=0.19\textheight]{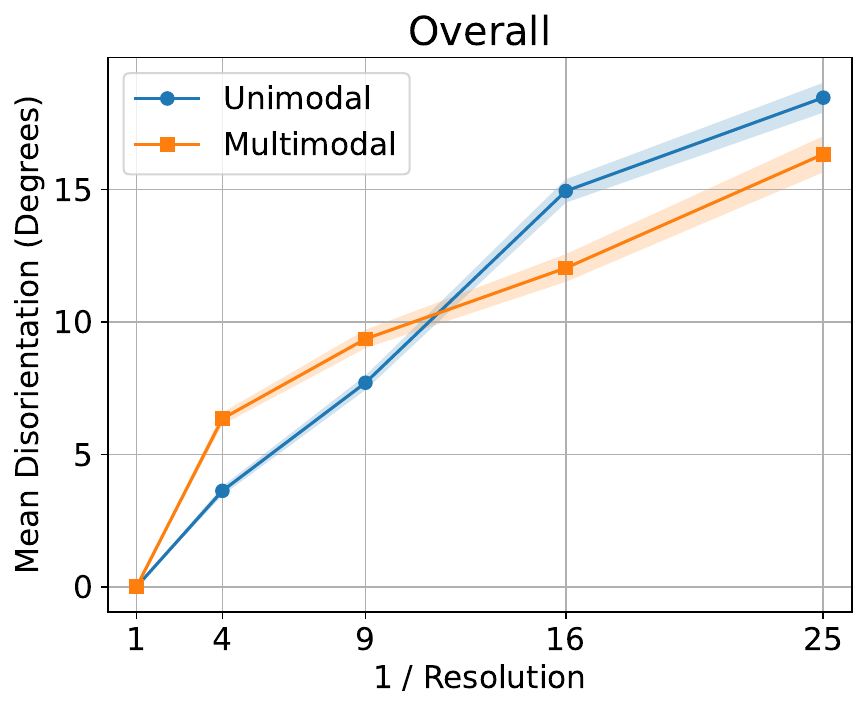}
\includegraphics[height=0.19\textheight]{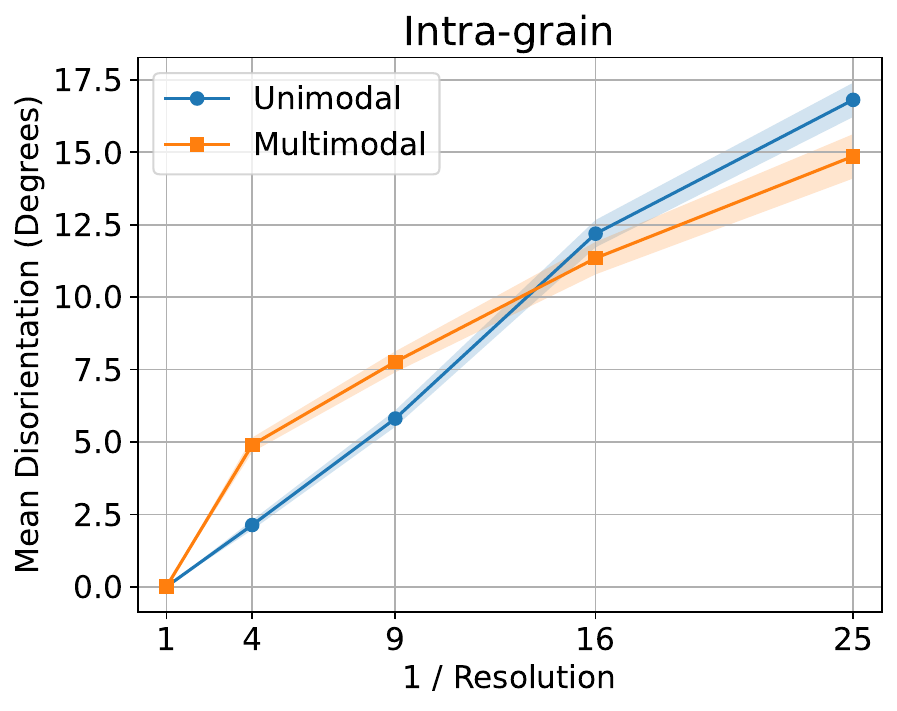}
\includegraphics[height=0.19\textheight]{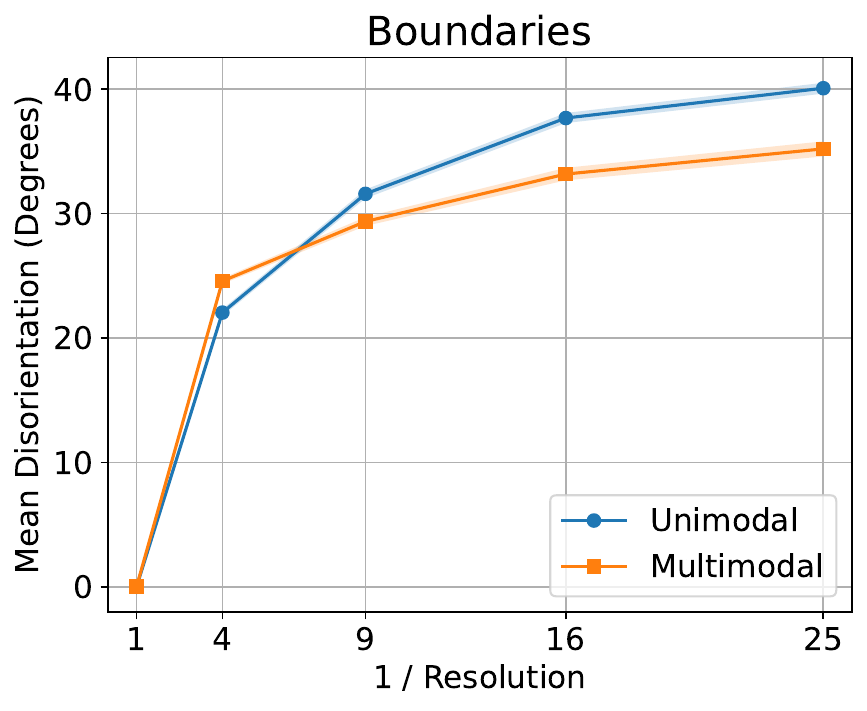}
\caption{Mean disorientation error across all unobserved EBSD pixels (left), unobserved intra-grain pixels (center), and unobserved boundary pixels (right) from varying super-resolution factors. Shaded regions indicate two standard errors. With lower resolution EBSD data, the multimodal diffusion model is more performant with improvements more concentrated along the grain boundaries.}
\label{fig:sr_vs_res}
\end{center}
\end{figure}

Our model also possesses strong performance on super-resolution. By scattering an observed resolution into the desired resolution and using this as the observed EBSD data, the models inpaint the remaining pixels. Shown in Figure~\ref{fig:sr_examples}, both the EBSD unimodal model $\cM_\text{E}$ and multimodal model $\cM_\text{EP}$ accurately perform EBSD super-resolution from $25\%$ of the pixels. As the observed resolution decreases, the recovery with the multimodal model produces crisper images than with the unimodal model. Quantitatively in Figure~\ref{fig:sr_vs_res}, we see that the multimodal model is able to produce more accurate EBSD images with lower resolution observations, as measured by disorientation \cite{larsen2017improved, ebsdtorch}, especially along the grain boundaries. This likely due to greater reliance on PL data for lower resolution images, making interpolation much more ill-posed for the unimodal diffusion model. The slight difference in quality between the diffusion models when we observe 1/4 resolution EBSD images may be related to model capacity issue discussed in Section~\ref{sec:no_multi}.

\subsubsection{Diffusion as a Denoiser}
\label{sec:denoise}

Finally, we take a qualitative look at an implicit perk of diffusion: denoising. By simply taking the average across all $N$ outputs from the inverse solver, we are able to significantly clean up white noise and scratches in PL data, as depicted in Figure~\ref{fig:denoise}. With more EBSD observations, the boundaries between grains with very similar PL measurements become more apparent.

\subsection{Error Analysis \& Limitations of All Methods}
\label{sec:error}

Looking closer at the outputs of these methods, we identify cases where some methods may be preferred over others.

\subsubsection{Noisy PL Data}

\begin{figure}[!t]
\begin{center}
\includegraphics[width=0.83\columnwidth]{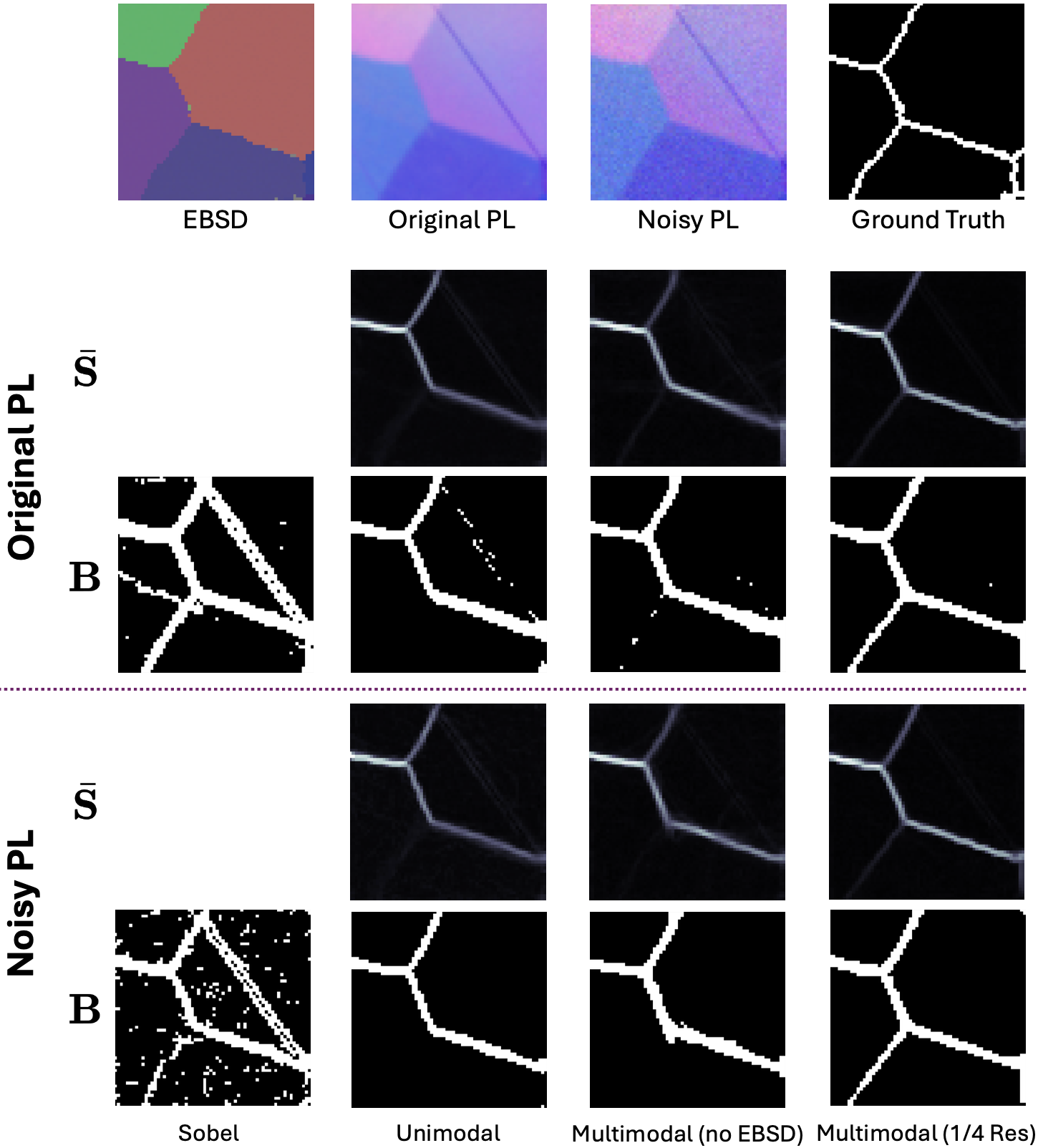}
\caption{The effect of noise in PL measurements. The diffusion models are more robust to noise and scratches compared to direct Sobel filtering. Noise is sampled from a centered Gaussian distribution with variance $0.05$. $\Bar{\bS}$ and $\bB$ are the aggregated Sobel maps and predicted boundaries, respectively.}
\label{fig:noisy_example}
\end{center}
\end{figure}

Direct Sobel filtering on PL data suffers in the presence of perturbations. In turn, this can affect boundary predictions. To highlight this issue, we inject zero-centered Gaussian noise with standard deviation $0.05$ into (already slightly noisy) observed PL data. Structured noise can also occur in real data, such as in the case of scratches, long thin streaks of discoloration. The effects of both of these types of noise are shown in Figure~\ref{fig:noisy_example}. In the case of Gaussian noise, the resulting Sobel output contains speckled predictions. For scratches, Sobel filtering incorrectly also labels the scratch as a boundary. In contrast, the diffusion models naturally perform denoising, dampening this issue, as we explored earlier in Section~\ref{sec:denoise} and Figure~\ref{fig:denoise}.

\subsubsection{Registration Error}

\begin{figure}[t]
\begin{center}
\includegraphics[width=\columnwidth]{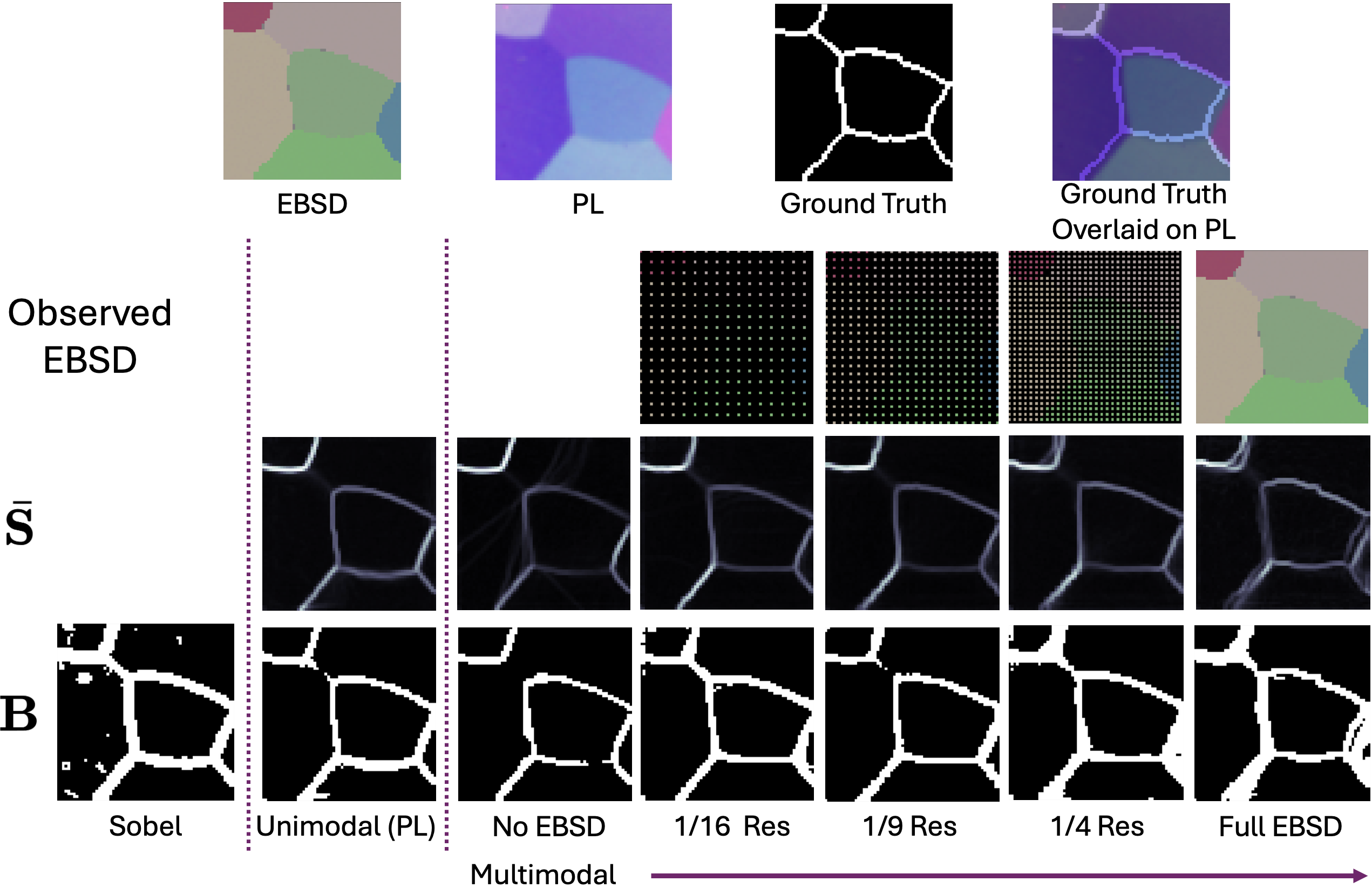}
\caption{Example where observing more EBSD harms the performance due to misalignment. From left to right, the top row shows the EBSD measurements, PL measurements, ground truth boundaries $\bB^\star$ derived from EBSD, and $\bB^\star$ overlaid on the PL image. The middle row shows the aggregated Sobel maps $\Bar{\bS}$ of diffusion-based methods with brighter colors indicating higher values. The bottom row shows the predicted boundaries $\bB$ of all methods. Misalignment between EBSD and PL causes noisy predictions in the form of thicker and dual boundaries.}
\label{fig:misaligned_example}
\end{center}
\end{figure}

Registering PL and EBSD data can result in slight errors (Figure~\ref{fig:labels}). While most pixels would be relatively unaffected due to the data's piecewise constant structure, disruptions to the boundaries can be significant. This is the reason behind using Gaussian blurred BCE loss and Chamfer loss as evaluation metrics. In addition to evaluation, registration error can also harm the generated reconstructions, too. Perhaps counterintuitively, observing more EBSD data can confuse the inverse solver if the two modalities have significant registration error. Shown in Figure~\ref{fig:misaligned_example}, when observing full EBSD data, grain boundary predictions are duplicated where one set corresponds to PL observations and the other shifted set corresponds to EBSD observations. With sparser EBSD observations, the grain boundaries in EBSD become less obvious, and therefore, duplicate boundaries become less of an issue. Interestingly, this raises another possible application of inverse solvers with multimodal diffusion for automatic image registration, which is a valuable future direction to explore.

\subsubsection{Subtle Grain Differences}

\begin{figure}[t]
\begin{center}
\includegraphics[width=\columnwidth]{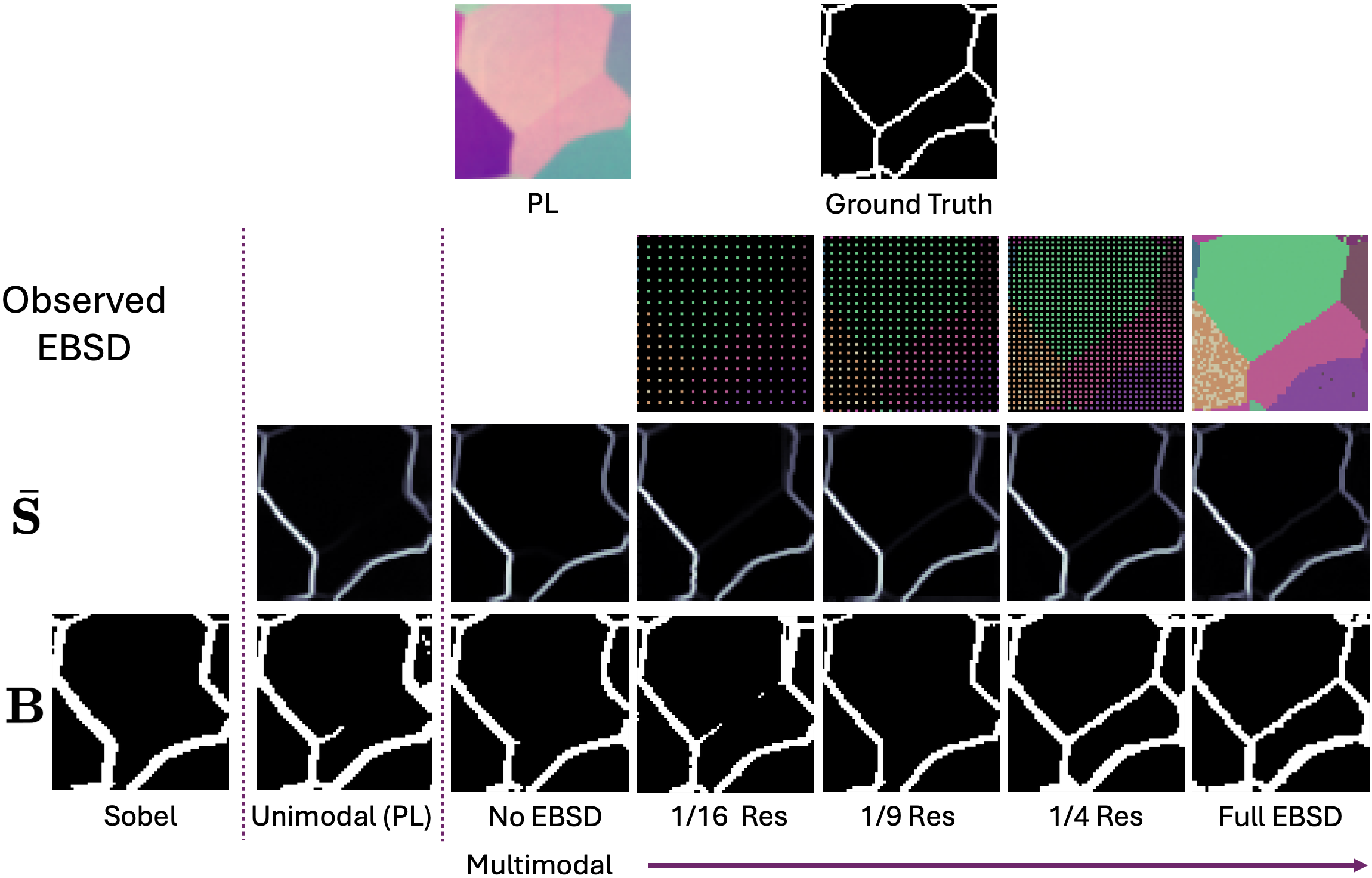}
\caption{Example where observing more EBSD provides clarity on similar PL values between grains. From left to right, the top row shows the PL measurements and ground truth boundaries $\bB^\star$ derived from EBSD. The second row shows the observed EBSD with increasing resolution from left to right. The third row shows the aggregated Sobel maps $\Bar{\bS}$ of diffusion-based methods with brighter colors indicating higher values (some pixels can be very faint). The bottom row shows the predicted boundaries $\bB$ of all methods. Best viewed zoomed in.}
\label{fig:gradual_example}
\end{center}
\end{figure}

Subtle changes between grains pose a challenge for all methods, though to varying extents. Depicted in Figure~\ref{fig:gradual_example} and third row of Figure~\ref{fig:denoise}, all methods have trouble separating neighboring grains with similar values. Fortunately, with the assistance of another modality, the inverse solvers with multimodal diffusion can more easily differentiate between individual grains.

\subsubsection{Missing Modality}\label{sec:no_multi}

With no observed EBSD, multimodal diffusion performs slightly worse than the unimodal diffusion model for boundary prediction (Figure~\ref{fig:boundary_metrics}). In fact, slight degradation when an entire modality is missing at inference for multimodal AI models is a phenomenon that has been observed beyond our setting and remains an active area of research \cite{wang2020makes, hagstrom2022adapt, deng2025words}. There are a couple hypotheses worth exploring in future work to investigate why this is happening. First is model capacity. While the unimodal diffusion model just needs to learn the structure of PL data, the multimodal diffusion model also needs to learn the structure of EBSD data and the interaction between EBSD and PL data within the same number training samples and parameters, making this a question of scale. Second is over-reliance on one modality \cite{wu2024deep, deng2025words}. Our multimodal diffusion model may be placing too much emphasis on EBSD data, posing a challenge for inverse solvers to properly guide the generation process when only PL observations are accessible.



\section{Conclusions}
\label{sec:conclusion}

Joint EBSD and PL multimodal diffusion shows immense promise as a base generative model applicable to numerous inverse problems where information from one or more modalities is needed. Trained solely on synthetic data, our method's efficacy transfers to real EBSD and PL data without additional training. Furthermore, by sampling multiple reconstructions with the same observation, we can obtain distributions of reconstruction which provide richer and higher quality content than a single deterministic prediction. Focusing particularly on grain boundary prediction, we show that low-resolution or sparse EBSD observations can greatly improve prediction quality. Along with demonstrations on denoising and super-resolution, multimodal diffusion is effective at integrating information from PL data to supplement low-resolution EBSD to accomplish these objectives. With particularly strong results at 1/4 resolution, this allows us to accelerate EBSD data collection by 4$\times$. There are still many exciting directions to explore to further push the capability of our models. For example, we would like to explore search-based algorithms for generation instead of just Monte Carlo sampling as we have done in this paper. Additionally, it would be interesting to examine the benefit of scaling the model and data which has seen success in traditional machine learning domains.

\section*{Acknowledgements}
The work of H.D., T.E., and Y.C. is supported in part by the Air Force D3OM2S Center of Excellence under FA8650-19-2-5209, by AFOSR under FA9550-25-1-0060, by ONR under N00014-19-1-2404, by NSF under ECCS-2126634, and by the Carnegie Mellon University Manufacturing Futures Initiative, made possible by the Richard King Mellon Foundation. The work of H.D. is also supported by the Wei Shen and Xuehong Zhang Presidential Fellowship at Carnegie Mellon University. T.E. is also supported by the National Science Foundation Graduate Research Fellowship under DGE2140739.
The authors thank Michael Uchic for valuable technical discussions and data collection.

\bibliographystyle{alphaabbr}
\bibliography{bibliography.bib}




\end{document}